\documentclass[aps,pra,twocolumn,notitlepage,superscriptaddress]{revtex4-1}
\usepackage{graphicx}  % needed for figures
\usepackage{dcolumn}   % needed for some tables
\usepackage{bm}        % for math
\usepackage{amssymb}   % for math
\usepackage{color}
\usepackage[table,dvipsnames]{xcolor}
\usepackage[colorlinks=true, linkcolor=blue, citecolor=blue,urlcolor=blue,breaklinks=true]{hyperref}

\usepackage{pifont}
\usepackage{amsmath, amsthm, amssymb}
\usepackage{amsfonts}
\usepackage{relsize}    % to resize the sum symbol etc
\usepackage{bbold}       % for identity in tensor notation
\usepackage{accents}    % \underaccent { accent } { symbol }

\usepackage{upgreek}
\usepackage{soul}
\newcommand{\beq}{\begin{equation}}
\newcommand{\eeq}{\end{equation}}
\newcommand{\bqa}{\begin{eqnarray}}
\newcommand{\eqa}{\end{eqnarray}}

\newcommand{\erf}[1]{Eq.~(\ref{#1})}

\newcommand{\trf}[1]{Table ~\ref{#1}} 

\newcommand{\frf}[1]{Fig.~\ref{#1}}

\newcommand{\dg}{^\dagger}

\definecolor{BLACK}{gray}{0}
\definecolor{RED}{rgb}{1,0,0}
\definecolor{GREEN}{rgb}{0.2,.6,0.2}
\definecolor{BLUE}{rgb}{0,0,1}

\renewcommand{\(}{\left(}
\renewcommand{\)}{\right)}
\newcommand{\sq}[1]{\left[ {#1} \right]}

\newcommand{\an}[1]{\left\langle{#1}\right\rangle}

\newcommand{\abs}[1]{\left| {#1} \right|}
\newcommand{\tr}[1]{{\rm Tr}\sq{ {#1} }}

\newcommand{\bra}[1]{\langle{#1}|}
\newcommand{\ket}[1]{|{#1}\rangle}
\newcommand{\ketbra}[2]{|{#1}\rangle\langle{#2}|} 
\newcommand{\ip}[2]{\langle{#1}|{#2}\rangle} 
\newcommand{\op}[2]{\ket{#1}\bra{#2}}

\renewcommand{\figurename}{{\bf Figure}}
\renewcommand{\thefigure}{{\bf \arabic{figure}}}

\begin{document}

\title{Experimental optical phase measurement approaching the exact Heisenberg limit}

\author{Shakib Daryanoosh}
\author{Sergei Slussarenko} %
\affiliation{Centre for Quantum Dynamics and Centre for Quantum Computation and Communication Technology, Griffith University, Brisbane, Queensland 4111, Australia}
\author{Dominic W. Berry} 
\affiliation{Department of Physics and Astronomy, Macquarie University, Sydney NSW 2113, Australia}
\author{Howard M.~Wiseman} \email{h.wiseman@griffith.edu.au}
\author{Geoff J. Pryde} \email{g.pryde@griffith.edu.au}
\affiliation{Centre for Quantum Dynamics and Centre for Quantum Computation and Communication Technology, Griffith University, Brisbane, Queensland 4111, Australia}

\begin{abstract}
\section*{Abstract}
The use of quantum resources can provide measurement precision beyond the shot-noise limit (SNL). The task of ab initio optical phase measurement---the estimation of a completely unknown phase---has been experimentally demonstrated with precision beyond the SNL, and even scaling like the ultimate bound, the Heisenberg limit (HL), but with an overhead factor. However, existing approaches have not been able---even in principle---to achieve the best possible precision, saturating the HL exactly. Here we demonstrate a scheme to achieve true HL phase measurement, using a combination of three techniques: entanglement, multiple samplings of the phase shift, and adaptive measurement. Our experimental demonstration of the scheme  uses two photonic qubits, one double passed, so that, for a successful coincidence detection, the number of photon-passes is $N=3$.  We achieve a precision that is within $4\%$ of the HL, surpassing the best precision theoretically achievable with simpler techniques with $N=3$. This work represents a fundamental achievement of the ultimate limits of metrology, and the scheme can be extended to higher $N$ and other physical systems.
\end{abstract}

\maketitle

\section*{Introduction}
Precise measurement is at the heart of science and technology~\cite{WisMil10}. An important fundamental concern is how to achieve the best precision in measuring a physical quantity, relative to the resources of the probe system. As physical resources are fundamentally quantised, it is quantum physics that determines the ultimate precision that can be achieved. Correlated quantum resources~\cite{GioMac06,Matthews17a,Matthews17b} such as entangled states can provide an enhancement over independent use of quantum systems in measurement.  

Quantum-enhanced optical phase estimation promises improvements in all measurement tasks for which interferometry is presently used~\cite{Cav81,GioMac11}. Such optical quantum metrology can be divided into two distinct tasks. In phase sensing, the goal is to determine small deviations in a phase about an already well-known value---a very specific situation. The use of maximally-path-entangled NOON states \cite{dowling08,NagTak07} can, in principle, provide optimal sensitivity for this task~\cite{slussarenko17n}. The more challenging task is phase measurement, sometimes called ab initio phase measurement~\cite{BerWis00}, in which the aim is to determine an unknown phase $\phi$ with no prior information about its value. In this case, the use of multiple passes of the optical phase shift and adaptive quantum measurement \cite{HigPry07}, or entanglement and adaptive quantum measurement \cite{XiaPry11}, have been shown to be capable of surpassing the shot-noise limit (SNL), $V^{\rm SNL} = 1/N$ (for large $N$). The SNL represents the minimum variance achievable with a definite number $N$ of independent samples of the phase shift by a photon. By making correlated samples of the phase shift, these schemes \cite{HigPry07,BerAnd15,XiaPry11} can achieve an asymptotic variance $V = ( B\pi/N)^2$. This is proportional to, but with a constant overhead $ B>1$ over, the ultimate limit (the Heisenberg limit, HL) of $(\pi/N)^2$ for the asymptotic ab initio task. 
To be precise, in terms of Holevo's variance measure \cite{holevo84,WisPry09}, the exact HL for any value of $N$ is 
\beq  \label{HL:var:gen}
V^{\rm HL} = \tan^2\left[\pi/\left(N+2\right)\right].
\eeq
Phase measurement schemes are not limited to optics: equivalent techniques have also used phase shifts of superposition states of single-NV-centre measurements induced by magnetic fields~\cite{waldherr12,nursan12}, for example.
\begin{figure*}
	\includegraphics{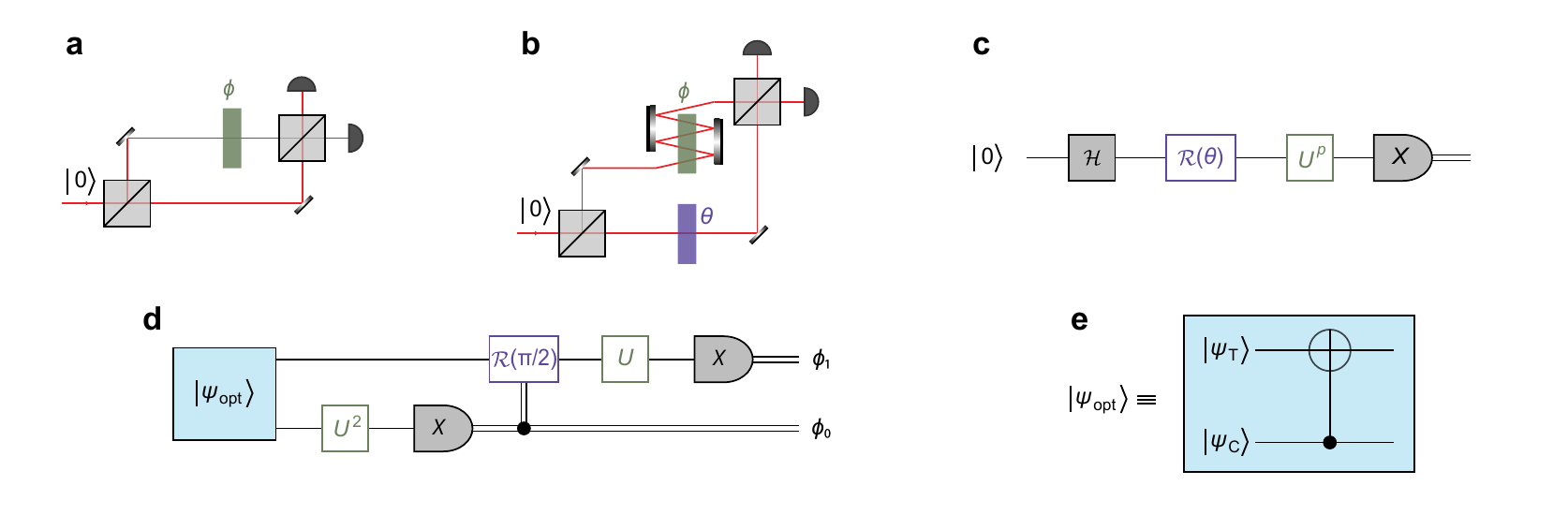}
	\caption{\label{fig:Qcircuit} {\bf Optical phase measurement concept.}  \textbf{a}, Basic  interferometric  setup for estimation of an unknown phase $\phi$.  \textbf{b}, Conceptual scheme of an advanced interferometer that includes multiple ($p$) passes of the phase shift $\phi$ and a controllable phase $\theta$ in the reference arm. {\bf c}, Quantum circuit representation of the interferometer shown in  \textbf{b}.  The interferometer is represented by a Hadamard gate $\mathcal{H}$ and a projective measurement in the $X$ basis, and the application of reference and unknown phases ($p$ passes) is represented by unitary operators $\mathcal{R}(\theta)$ and $U^p$, respectively.  {\bf d},  Quantum circuit for Heisenberg-limited interferometric phase estimation with $N=3$ resources. The protocol is extensible to higher $N$, in principle~\cite{WisPry09}. {\bf e}, Quantum circuit for the preparation of the optimal state $\ket{\psi_{\rm opt}}$, \erf{ket_opt}, using a CNOT gate with control and target qubits prepared in $\ket{\psi_{\rm C}}$ and $\ket{\psi_{\rm T}}$, respectively.}
\end{figure*}
Here we demonstrate a technique to address this outstanding, fundamental question of quantum metrology: how to measure phase at the exact HL? We show a concrete way to implement the conceptual scheme previously proposed in theory \cite{WisPry09}, and implement it experimentally.  As in previous photonic ab initio phase estimation experiments, we characterise the quality of our implementation with respect to detected resources -- it relies on probabilistic state preparation and measurement schemes, and takes into account only the successful coincidence detections in the calculation of precision. We thus prove the principle of the scheme, which in future can be extended to remove postselective elements.
\section*{Results}
\subsection*{Theory}
We begin by introducing the basic tools and techniques used in this work.  The basic concept of optical phase measurement with photons is shown in \frf{fig:Qcircuit}\textbf{a}. The phase to be measured is inserted in one path of an interferometer; the other path is the reference arm. 
In the language of quantum information, a photon incident on the first beam splitter (BS) 
is represented by the logical state $\ket{0}$. The action of the BS is modelled by a Hadamard gate $ {\rm {\mathcal H}} \ket{0} = \(\ket{0} + \ket{1}\) / \sqrt{2}$. The unknown phase shift applied on the path representing $\ket{1}$ is implemented by the unitary gate $U(\phi) = \exp(i\phi\op{1}{1})$. The last BS prior to detection stages maps the logical $Z$ basis onto the $X$ basis. 

A more general protocol may include more sophisticated techniques. The relevant constituents are: the quantum state of the light in the interferometer paths; the possibility of multiple coherent samplings of the phase shift by some photons; and the detection strategy. For example, \frf{fig:Qcircuit}\textbf{b} generalises the basic single photon interferometer to include $p\geq1$ applications of $U(\phi)$ and a classically controllable phase, described by ${\rm \mathcal{R}}(\theta) = \exp(i\theta \op{0}{0})$, on the reference path (representing $\ket{0}$). We can also depict  this interferometer following the quantum circuit convention, as in \frf{fig:Qcircuit}\textbf{c}.

For ab initio phase measurement with $N$ photons and no multipassing ($p=1$), it is known theoretically that the HL can be achieved by preparing a path-entangled state~\cite{BerWis00,BerBre01} and implementing an entangling detection scheme~\cite{sanders95}. The problem is that both of these steps are very difficult to do. An alternative way~\cite{WisPry09} to achieve the HL uses entanglement across multiple spatio-temporal modes, and multiple applications $p$ of the phase gate, combined with the inverse quantum Fourier transform (IQFT) for the measurement. While the IQFT is also an entangling operation, it has been known for some time~\cite{GriNiu96} that, in this phase estimation algorithm (PEA)~\cite{NieChu01}, it can be replaced by an adaptive measurement scheme~\cite{WisMil10}, where individual photons are measured one by one, with the reference phase adjusted after each measurement. This replacement requires the photons in the entangled state to be spread out in time, but suffers no penalty in measurement precision. 

Here, we show the practicality of combining entanglement, multipassing and adaptive measurement to achieve the HL. Our Heisenberg-limited interferometric phase estimation algorithm (HPEA) \cite{WisPry09} is illustrated in \frf{fig:Qcircuit}\textbf{d}. This protocol is based on the standard PEA such that using $K+1$ qubits yields an estimate $\phi_{\rm est}$ of the true phase $\phi$ with $K+1$ bits of precision~\cite{NieChu01}. It involves application of the phase gate $N=2^{K+1}-1$ times, with the number of applications being $p = 2^K, 2^{K-1}, \cdots, 2^0$ on each successive qubit (photon). Our particular demonstration is an instance of a $(K+1=)$ 2-photon superposition state \cite{WisPry09} that may be used to perform a protocol with $N=2^{K+1}-1=3$ resources, achieving a variance for ab initio phase estimation of exactly $V^{\rm HL}$, \erf{HL:var:gen}. 

The optimal entangled state for the HPEA is~\cite{WisPry09}
\beq  \label{ket_opt}
\ket{\psi_{\rm opt}} = c_0 \ket{\Phi^+} + c_1 \ket{\Psi^+}, 
\eeq 
where 
\beq
c_j = \frac{\sin\left[(j+1)\pi/5\right]}{\sqrt{\sum_{k=0}^1 \sin^2 \left[(k+1)\pi/5\right]}}, 
\eeq
and where $\ket{\Phi^+}=\left(\ket{00}+\ket{11}\right)/\sqrt{2}$ and $\ket{\Psi^+}=\left(\ket{01}+\ket{10}\right)/\sqrt{2}$ are Bell states. The optimal adaptive measurement~\cite{GriNiu96} is implemented by measuring the qubits sequentially in the $X$ basis, and, conditioned on the results, adjusting the controllable phase $\theta$ shifts on subsequent qubits, as shown in \frf{fig:Qcircuit}\textbf{d}. 
\begin{figure}
	\includegraphics{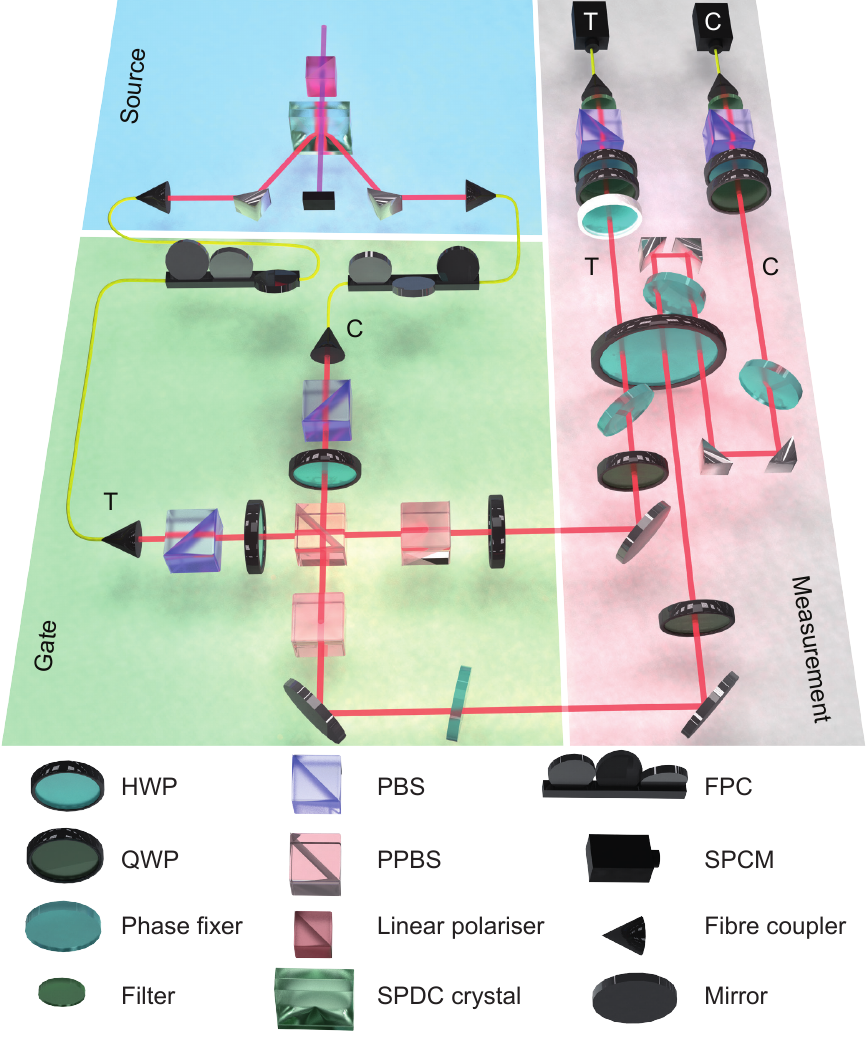}
	\caption{\label{fig:setup}  {\bf Schematic of the experimental setup.} Single photons at $820~{\rm nm}$ are generated via a type-I spontaneous parametric downconversion (SPDC) process (blue background) and collected using single-mode fibres and passed into the entangling gate (green background) in order to realise the state $\ket{\psi_{\rm exp}}$. Input polarisation was adjusted with fibre polarisation controllers (FPC). The nondeterministic universal CNOT gate, composed of 3 partially polarising beam splitters (PPBS) and 2 half-wave plates (HWP), performs the state preparation by post-selecting coincidence events between the control and target output ports with success probability $1/9$. The area with grey background corresponds to the implementation of the phase estimation. Photons in mode C pass twice through the HWP (acting as a phase shift element), in order to realise the $U(\phi)^2$ operation. Photons in mode T experience the phase shift once (performing the $U(\phi)$ operation). The effect of the feedforward operation, $\mathcal{R}(\theta)$, is simulated by dialling a HWP (depicted with a white rim), for a fixed time period, with $0$ and $\pi/8$ corresponding to the ON and OFF settings of the control operation. Finally, photons are independently directed to a polarisation analysis unit consisting of a quarter-wave plate (QWP), HWP and a polarising beam splitter (PBS) followed by a $2~\rm{nm}$ spectral filter and a single photon counting module (SPCM). See Methods for further details on the experimental setup operation.}
\end{figure}
\subsection*{Experimental scheme}
\begin{figure}[b]
	\includegraphics{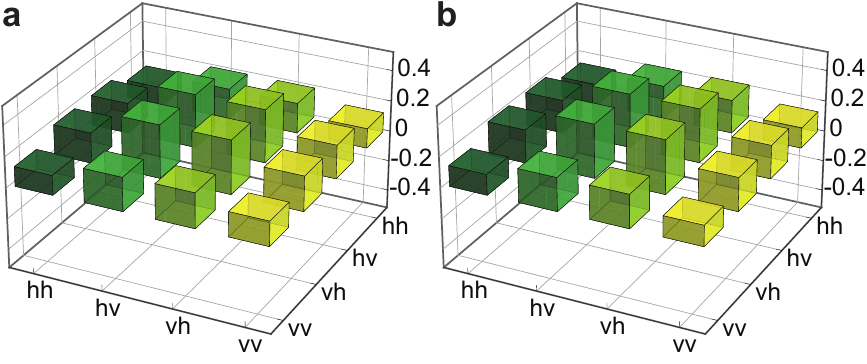}
	\caption{\label{fig:tomo}{\bf Density matrices of the experimental state and $\bm{\rho_{\rm opt}}$.} {\bf a}, Real part of the state matrix $\rho_{\rm exp}$ reconstructed with polarisation state tomography. The fidelity of the state with the optimal state $\ket{\psi_{\rm opt}}$, \erf{ket_opt}, is $\bra{\psi_{\rm opt}}\rho_{\rm exp}\ket{\psi_{\rm opt}} = 0.980\pm0.003$, and the purity is $\tr{\rho^2_{\rm exp}}= 0.965\pm0.006$. The density matrix was calculated from approximately $50,000$ two-fold coincidence events. Uncertainties in fidelity and purity represent $95\%$ confidence intervals calculated with Monte-Carlo simulation~\cite{white07}. Imaginary components (not shown) are $\leq0.013$. {\bf b}, Real part of the ideal optimal state $\rho_{\rm opt}$.  Note that  $\textrm{Im}(\rho_\textrm{opt})=0$.}
\end{figure}

In our experiment (\frf{fig:setup}), we used orthogonal right- and left-circular polarisations instead of paths to form the two arms of the interferometer. We used a nondeterministic CNOT gate, acting on photon polarisation qubits (horizontal $\ket{{\rm h}}\equiv\ket{0}$, vertical $\ket{{\rm v}}\equiv\ket{1}$), to generate the state in Eq.~(\ref{ket_opt}). As shown in \frf{fig:Qcircuit}\textbf{e}, the control qubit is prepared in the diagonal polarisation state $\ket{\psi_{\rm C}} = \(\ket{{\rm h}} + \ket{{\rm v}}\)/\sqrt{2}$, and the target qubit in the linear polarisation $\ket{\psi_{\rm T}} = c_0 \ket{{\rm h}} + c_1 \ket{{\rm v}}$, so that the output state after the CNOT is the optimal state: $\ket{\psi_{\rm opt}} = \hat{U}_{\rm CNOT} \(\ket{\psi_{\rm C}} \otimes \ket{\psi_{\rm T}}\)$. Figure \ref{fig:tomo} shows the density matrices of the experimentally generated state $\rho_{\rm exp}$ and the ideal state $\rho_{\rm opt} \equiv\ketbra{\psi_{\rm opt}}{\psi_{\rm opt}}$. 

The polarisation interferometer, highlighted by the grey background in \frf{fig:setup}, used a large half-wave plate (HWP) to implement the unknown phase shift between the arms. Mode ${\rm C}$ was passed twice through this unknown phase. Another HWP (shown in \frf{fig:setup} with a white rim) was used as the reference phase shift $\theta$ on mode ${\rm T}$, in order to implement the detection scheme.
We implemented the feedforward step nondeterministically, using waveplates that were fixed for each run, combined with postselective sorting of the data based on the results from the detector labeled ${\rm C}$. Although this approach would be inadequate for estimation from exactly one shot, it is an accurate  way to characterise the performance of the scheme over many repetitions. \trf{firing:pat} shows how the data was sorted and how phase values were allocated for each shot, according to the detector firing patterns. 
\begin{table}[h]
	\caption{\label{firing:pat} The detection outcome patterns.}
	\begin{tabular}{ c c c c c}
		\hline \hline
		\begin{tabular}{@{}c@{}}Outcome in\\${\rm C}$\;\;\end{tabular} & \begin{tabular}{@{}c@{}}\;\;$\theta$ \;\; \end{tabular} &  \begin{tabular}{@{}c@{}} 
			Successful events \\in ${\rm CT}$ \end{tabular} & & \begin{tabular}{@{}c@{}} Rejected events\\in ${\rm CT}$   \end{tabular}\\[8pt] \hline
		${\rm d}$ & 0 & ${\rm dd},{\rm da}$ & & ${\rm ad},{\rm aa}$\\[4pt] 
		${\rm a}$ & $\pi/8$ & ${\rm ad},{\rm aa}$ & & ${\rm dd},{\rm da}$  \\ 
		\hline \hline
	\end{tabular}
\end{table}

\subsection*{Experimental phase estimation}
\begin{figure*}[t]
	\includegraphics{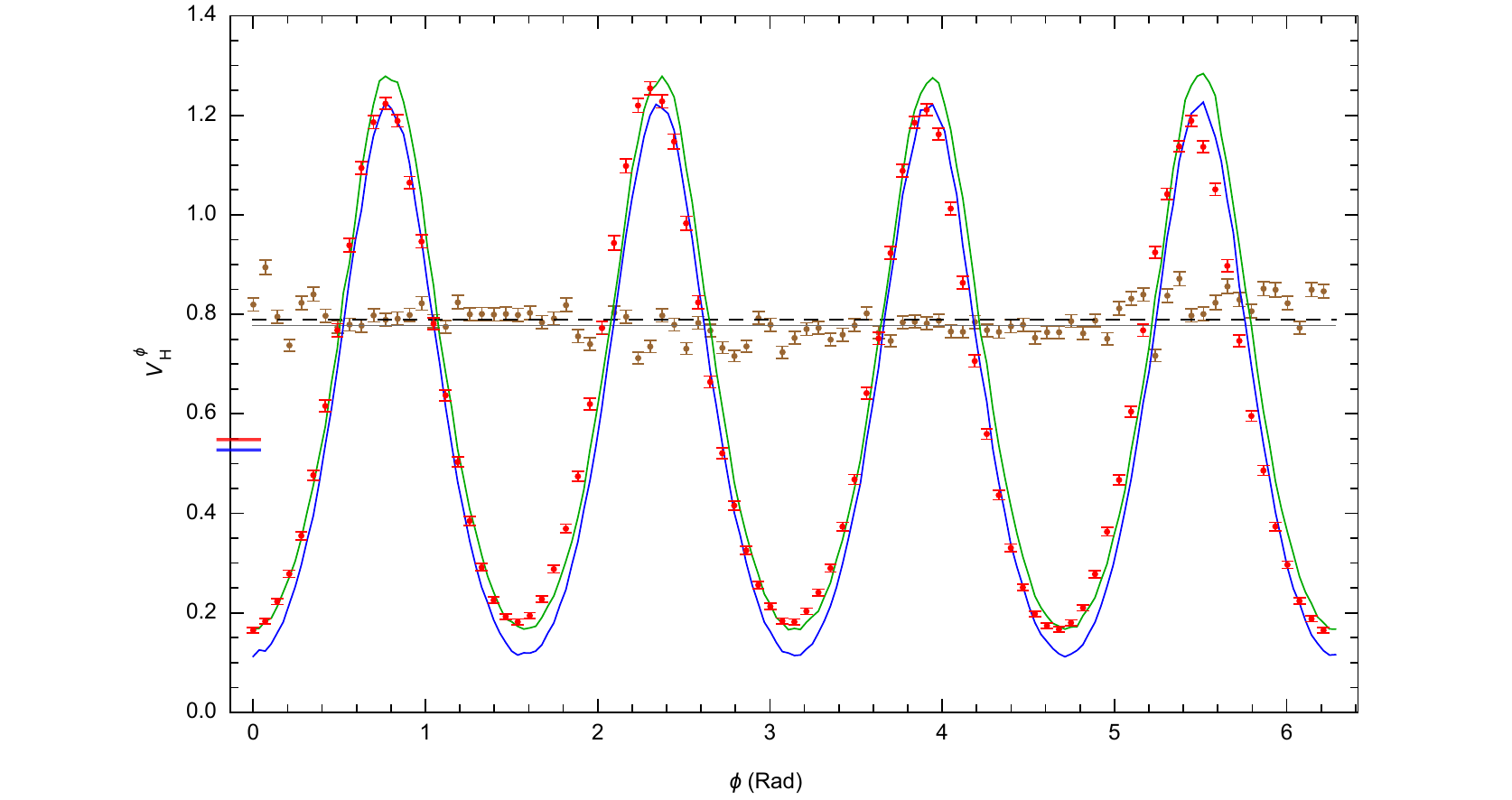}
	\caption{\label{fig:holvar}{\bf Heisenberg-limited phase estimation with $\bm{N=3}$ resources}.  Red dots represent experimentally measured variance $V_{\rm H}^\phi$ as a function of $\phi$.  The red horizontal line-segment cutting the left axis shows the optimal  protocol Holevo variance $V_{\rm H}=0.5497 \pm 0.0007$, determined from these data, while blue line-segment shows the HL. The blue and the green curves represent results of numerical simulations of the variance for the ideal optimal state $\rho_{\rm opt}$ and experimentally prepared state $\rho_{\rm exp}$, respectively. Brown dots represent $V_{\rm H}^\phi$ for the shot-noise-limited interferometry and the black dashed line represents the measured Holevo variance $V_{\rm H}=0.7870\pm 0.0007$ for the same measurement. The grey solid line shows the SNL. Numerical values for the experimental results and corresponding limits are detailed in \trf{tbl:HV}.  Each data point was calculated from at least $50,000$  two-fold coincidence events and the error bars represent $95\%$ confidence intervals calculated with the bootstrap method \cite{DavHin97}}
\end{figure*}

To characterise the performance of our HPEA, we first calculate the conditional Holevo variance $V_{\rm H}^\phi$ in the estimates for each applied phase $\phi$ (see Methods for details on data analysis). Here $V_{\rm H}^\phi = |\left\langle{\exp}[i(\phi - \phi_{\rm est})]\right\rangle_{\phi_{\rm est}}|^{-2}-1$ for a given $\phi$, where $\left\langle\dots\right\rangle_{\phi_{\rm est}}$ indicates averaging over the values of $\phi_{\rm est}$ resulting from the data. Figure~\ref{fig:holvar} shows $V_{\rm H}^\phi$ for the entire range of $\phi \in [0,2\pi)$. The protocol performs best when $\phi = 0,\pi/2,\pi$, and $3\pi/2$, corresponding to the cases where, to a good approximation, only one of the four possible detection outcomes occur:  ${\rm dd}$, ${\rm ad}$, ${\rm da}$, and ${\rm aa}$,  respectively, as shown in \frf{fig:prob}. (Here,  ${\rm d}$(${\rm a}$)  means the diagonal (antidiagonal) polarisation states, which are $X$-basis eigenstates.) It performs worst for intermediate phases. This explains the oscillatory nature of the data in \frf{fig:holvar}. 
Since we are interested in evaluating the precision of ab initio phase estimation, we cannot use any knowledge of $\phi$. 
Thus we erase any initial phase information by calculating the unconditional Holevo variance  
$V_{\rm H}=|\langle\langle{\exp}[i(\phi - \phi_{\rm est})]\rangle_{\phi_{\rm est}}\rangle_{\phi}|^{-2}-1$, which averages over $\phi$. We find $V_{\rm H}=0.5497 \pm 0.0007$, whereas the Heisenberg limit for $N=3$ resources  is $V^{\rm HL} \approx 0.5278$~\cite{BerWis09}.  As can be seen from the simulation  (described in  Appendix~\ref{appnD})  results in \frf{fig:holvar},  this { 4\%}  discrepancy between the experimental result and theoretical bound  can be  attributed to the non-unit fidelity of the prepared entangled state with respect to $\rho_{\rm opt}$, highlighting the strong correlation between the protocol performance and quality of the prepared state~\cite{modi2016}.  The small phase offset between the measured data and numerical simulations appears due to a residual phase shift from mirrors and other optical components. This constant phase offset does not influence HPEA precision and can be compensated by a more sophisticated calibration of the setup, or in post-processing, if required. 

For comparison, we perform standard quantum interferometry with three independent photons (see  Appendices~\ref{appnE1} and~\ref{appnE2} for details). Calculating the Holevo variance for this measurement gives $V_{\rm H} = 0.7870\pm0.0007$ which is close to the theoretical value of $V^{\rm SLN}=0.7778$ for the SNL with $N=3$ resources.
\begin{figure}[h]
	\includegraphics{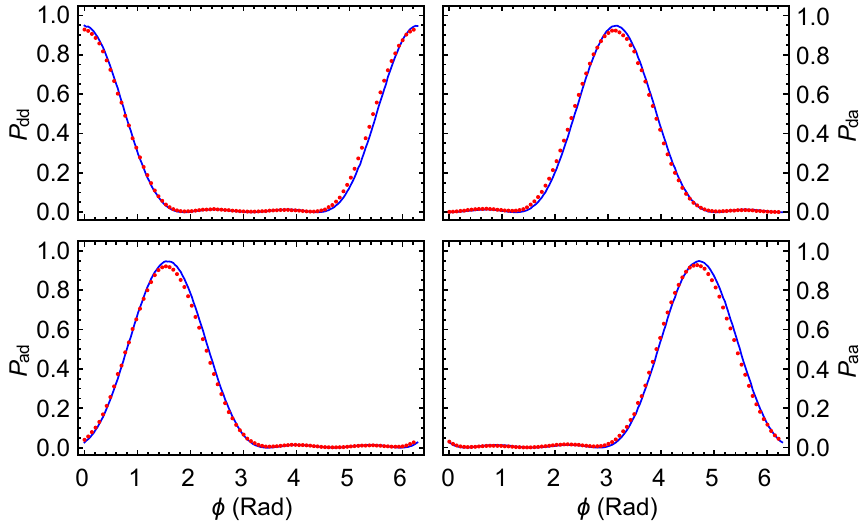}
	\caption{\label{fig:prob} {\bf Probability distribution of measurement outcomes}. The   probabilities of obtaining the four possible $\{{\rm dd},{\rm ad},{\rm da},{\rm aa}\}$  measurement outcomes which correspond to four possible $\phi_{\rm est}$ values, for each phase value shown in \frf{fig:holvar}. The variance $V^\phi_{\rm H}$ is minimised for those $\phi$ values when one of the probabilities is maximum. Dots are experimental values and lines are numerical simulations that use the experimentally generated $\rho_{\rm exp}$ as input. Error bars, representing the statistical uncertainty due to the finite number of
		measurement sets,  are smaller than the dot size.}
\end{figure}
\begin{table}[h]
	\caption{\label{tbl:HV} The Holevo variance for different schemes.}
	\begin{tabular}{ c  c  c  c }
		\hline \hline
		\begin{tabular}{@{}c@{}}Symmetric\\entanglement\;\end{tabular} & \begin{tabular}{@{}c@{}}Multipass \; \end{tabular} &  \begin{tabular}{@{}c@{}} Adaptive \\ measurement \end{tabular} & $V_{\rm H}$ \\ \hline
		\ding{51} &\ding{51} & \ding{51} & $0.5278$    \\ 
		\ding{51} &\ding{51} & \ding{51} & $0.5497(7)$ (Exp.)  \\ 
		\ding{51} &\ding{55} & \ding{51} & $0.5569$ \cite{BerryTh}   \\ 
		\ding{55} &\ding{51} & \ding{51} & $0.5609$ \\ 
		\ding{51} &\ding{51} & \ding{55} & $0.6547$ \\ 
		\ding{55} &\ding{55} & \ding{55} & $0.7778$    \\ 
		\ding{55} &\ding{55} & \ding{55} & $0.7870(7)$ (Exp.)  \\ 
		\hline \hline
	\end{tabular}
\end{table}

We also compare our results with the theoretically-optimal results for other schemes that use a subset of the three protocol components; see \trf{tbl:HV}. It can readily be observed that our scheme outperforms all those that use two of the components only.  While the experimentally measured $V_{\rm H}$ is numerically only a little lower than the next best theoretical bound (see Appendix~\ref{appnF}  for derivation of theory results), the difference amounts to a $10$ standard deviation improvement. We note that arbitrary entanglement can always do the job of multiple passes, by replacing each multipassed photon with a multiple-photon NOON state~\cite{dowling08}, split across the two polarisations. Thus our results could, in principle, be reproduced by an entangled state of three photonic qubits, two in one spatio-temporal mode and the third in another, with both modes going through $U(\phi)$ once. We rule out such complicated schemes in our comparison by restricting to symmetric entanglement, in which each photon that passes through $U(\phi)$ a given number of times is prepared identically. (This is the case for the entanglement in our scheme since each of the two photons passes through $U(\phi)$ a different number of times.) 

\section*{Discussion}
We have experimentally demonstrated how to use entanglement, adaptive measurement and multiple passes of the phase shift to perform ab initio phase measurement that outperforms any other scheme, in terms of sensitivity per resource. Our results are very close to the Heisenberg limit for $N=3$, giving substantial experimental justification to the theoretical prediction that this method can  achieve  the ultimate measurement sensitivity.  While in our analysis we count only photons  detected, in two-fold coincidences consistent with success of the probabilistic operations,  as resources, advances in nascent photon source~\cite{weston16} and detection~\cite{marsili13} technology, heralded state preparation schemes~\cite{barz10,ulanov16} and deterministic adaptive measurement (with \emph{e.g.}\ a Pockels cell) may soon allow saturation of the Heisenberg limit bound even when all the employed resources are taken into account. Since quantum phase-sensitive states are susceptible to loss~\cite{knysh11}, we expect that similar considerations would apply to the states in our scheme. For small $N$, as we use here, loss has less of an effect on the sensitivity.  Future extensions to the scheme will employ $K+1>2$ photons, yielding $N=2^{K+1}-1$ resources and a correspondingly decreased phase uncertainty, as quantum logic circuits become increasingly capable of producing large entangled states with high fidelity. We note that while we have implemented this scheme optically, it can be applied to the estimation of any parameter that implements a phase shift between qubit states of some physical system.

\section*{Methods}
\subsection*{Photon source} \label{appnA} 
We used spontaneous parametric downconversion (SPDC) to produce pairs of polarisation-unentangled single photons. Ultrashort pulses from a mode-locked Ti:sapphire laser at $820\,{\rm nm}$ with repetition rate of $80\,{\rm MHz}$, were upconverted to  $410\,{\rm nm}$ wavelength through a second-harmonic-generation (SHG) process with a $2\,{\rm mm}$ lithium triborate (LBO) crystal. The SHG beam was collimated with a $f=75\,{\rm mm}$ lens and the IR pump was spatially filtered away with two dispersive prisms. The UV light was focused on a $0.5\,{\rm mm}$ BiBO crystal to generate photon pairs via a type-I SPDC. The pump power was set to approximately $100\,{\rm mW}$ to ensure low probability of double pair emission from the crystal. Using $2\,{\rm nm}$ narrowband spectral filters, and Excelitas single photon counting modules (SPCMs) with detection efficiency in the range $(50-60)\%$, the  overall  coincidence efficiency was in the window of $(11-13)\%$ with single-detection count rates of $\sim 40\,{\rm KHz}$.

\subsection*{Entangling gate} \label{appnB} 
The single photons produced in the SPDC process were spatially filtered using antireflection (AR) coated single-mode fibres, and sent through the entangling gate to produce a state close to the optimal state $\rho_{\rm opt}$. The logical circuit of the gate consisted of three PPBSs, with $\eta_{\rm v}=1/3$ and $\eta_{\rm h}=1$ for the transmissivity of vertically and horizontally polarised light respectively, to produce a nondeterministic controlled-$Z$ operation~\cite{Ralph2009}.  Two of the PPBSs  were oriented $90^\circ$ (around the photon propagation axis) such that $\eta_{\rm v}=1$ and $\eta_{\rm h}=1/3$, as illustrated in \frf{fig:setup}.  Two HWPs oriented at $22.5^\circ$ with respect to the optical axis were used to perform the Hadamard operations required for the correct operation of the CNOT gate.  The successful operation of the gate is heralded by the presence of one photon in each output mode of the gate, with overall success probability of $1/9$. At the core of this realisation is the nonclassical interference that occurs between vertically polarised photons in modes C and T impinging on the central PPBS, see \frf{fig:setup}. The maximum interference visibility that can be observed with  $\eta_{\rm v}=1/3$ transmissivity is  $0.8$. We observed  $0.790\pm0.005$  visibility (see Appendix \frf{fig:hom}) Hong-Ou-Mandel interference~\cite{HOM87}, indicating excellent performance of the gate.  In the measurement with three uncorrelated resources, input photon polarisations were set to $\ket{{\rm h}}$, so the photons propagated through the gate without undergoing nonclassical interference, but still suffering $2/3$ loss in each mode. Photons in mode C were sent to a SPCM and acted as heralds for photons in mode T, which in turn were used to perform the shot-noise-limited interferometry.

\subsection*{ Phase shifts and probabilistic adaptive measurements} \label{appnC} 
To encode both unknown and classically controllable phases we proceeded as follows. The prepared state at the end of the entangling gate is ideally in the form of $\ket{\psi_{\rm opt}} = c_0 \ket{\Phi^+} + c_1 \ket{\Psi^+}$, \erf{ket_opt}, which is a superposition of the Bell states, $\ket{\Phi^+} = (\ket{{\rm hh}}+\ket{{\rm vv}})/\sqrt{2}$, and $\ket{\Psi^+} = (\ket{{\rm hv}}+\ket{{\rm vh}})/\sqrt{2}$. Here ${\rm h}$ and ${\rm v}$ are horizontal and vertical, respectively, polarisation states of a single photon, and encode the logical $\ket{0}$ and $\ket{1}$ states of a qubit. The linear polarisations were transformed to circular ones prior to the application of the phase shift. This was done by a QWP set at $\pi/4$, yielding
\beq  \label{LinToCir}
\begin{pmatrix}
	\ket{{\rm h}}\\
	\ket{{\rm v}}
\end{pmatrix} \xrightarrow{U_{\rm Q}^{(\pi/4)}}
\begin{pmatrix}
	e^{i \pi/4} \ket{{\rm r}} \\
	e^{-i \pi/4} \ket{{\rm l}}
\end{pmatrix}. 
\eeq
Here $U_{\rm Q}^{(\gamma)}$ is the unitary operation for a QWP  with optic axis oriented at $\gamma$ with respect to horizontal axis. The phase shift of $\phi$ between the right (${\rm r}$) and left (${\rm l}$) circular polarisations could then be applied by setting the 2-inch HWP in \frf{fig:setup} at $-\phi/4+\pi/8$, producing the transformation
\beq  \label{CirToPhi}
\begin{pmatrix}
	e^{i \pi/4} \ket{{\rm r}} \\
	e^{-i \pi/4} \ket{{\rm l}}
\end{pmatrix} \xrightarrow{U_{\rm H}^{(-\phi/4+\pi/8)}}
\begin{pmatrix}
	e^{i\phi} \ket{{\rm l}} \\
	\ket{{\rm r}}
\end{pmatrix},
\eeq
where we have ignored the global phase factor, and $U_{\rm H}^{(\gamma)}$ is the  operator  of a HWP with optic axis set at $\gamma$. We implemented the feedforward operation through the same procedure. By analogy with \eqref{LinToCir} and \eqref{CirToPhi}, implementing the feedforward operation by itself, setting the corresponding HWP at $\theta/4+\pi/8$, gives
\beq  \label{LinToCirToTheta}
\begin{pmatrix}
	\ket{{\rm h}}\\
	\ket{{\rm v}}
\end{pmatrix} \xrightarrow{U_{\rm Q}^{(\pi/4)}}
\begin{pmatrix}
	e^{i \pi/4} \ket{{\rm r}} \\
	e^{-i \pi/4} \ket{{\rm l}}
\end{pmatrix}  \xrightarrow{U_{\rm H}^{(\theta/4+\pi/8)}}
\begin{pmatrix}
	\ket{{\rm l}} \\
	e^{i\theta} \ket{{\rm r}}
\end{pmatrix}.
\eeq
Combining both allowed us to encode the phase shift $\phi-\theta$ between the two arms of the interferometer. 

The next step was to perform the adaptive measurements, which we implemented in a probabilistic manner. Since the feedback-controlled unitary operation $\mathcal{R}(\theta)$ has only two settings in this scheme, we set the corresponding HWP at $\theta=0$ and collected data for a fixed period of time. We recorded only those coincidence events where detector ${\rm C}$ (\frf{fig:setup}) registered a ${\rm d}$-polarised photon, as shown in \trf{firing:pat}. We repeated this for $\theta=\pi/8$ and detection of ${\rm a}$ polarisation at detector ${\rm C}$. 
In other words, when the photon in mode ${\rm C}$ is projected onto $\ket{{\rm d}}$ ($\ket{{\rm a}}$)  state, it is expected that the feedforward unit is in an OFF (ON) setting, equivalent to dialling $\theta = 0\, (\theta = \pi/8)$ for the HWP acting on the photon in mode ${\rm T}$. This provides for characterisation of the protocol performance without active switching. 

Each single shot detection (recorded coincidence) provides  $\phi_{\rm est} = \pi(\phi_0 \times 2^{0}+\phi_1 \times 2^{1})/2$. Here, $\phi_0\phi_1 \in \{00,01,10,11\} \leftrightarrow \{{\rm dd},{\rm ad},{\rm da},{\rm aa}\}$. The probability of obtaining the $\phi_0 \phi_1$ result is equal to the number of times $n_{\phi_0 \phi_1}$ that this measurement result occurs,  divided by the size of the ensemble $n_{\rm ens}$ over which the Holevo variance is calculated. Thus from the measurement record we evaluated the true phase $\phi$ using the relation
\beq \label{TruePhaseCal}
\phi \approx {\rm arg} \left[\frac{1}{n_{\rm ens}} \;\sum_{\phi_0=0}^1\; \sum_{\phi_1=0}^1 \; n_{\phi_0 \phi_1} \;{\rm exp}\({i \phi_{\rm est}}\)\right],
\eeq 
which becomes exact when $n_{\rm ens}\rightarrow\infty$. The conditional Holevo variance $V_{\rm H}^\phi$ is then calculated according to $V_{\rm H}^\phi= |\an{s}_{\phi_{\rm est}}|^{-2}-1$, with $s = {\rm exp}\left[i\(\phi-\phi_{\rm est}\)\right]$. Finally, the unconditional Holevo variance~\cite{BerBre01,BerWis09} is calculated as $V_{\rm H}= |\an{s}_{\phi_{\rm est},{\rm \phi}}|^{-2}-1$,  or, equivalently, 
\beq \label{hol_via_nonhol}
V_{\rm H}= \left|\an{(V_{\rm H}^\phi+1)^{-1/2}}_\phi\right|^{-2}-1. 
\eeq

\noindent\textbf{Acknowledgments}

\noindent The authors thank R.\ B.\ Patel for assistance with data acquisition code. This research was supported by the Australian Research Council Centre of Excellence Grant No.\ CE110001027. S.D.\ acknowledges financial support through an Australian Government Research Training Program Scholarship. D.W.B.\ is funded by an Australian Research Council Discovery Project Grant No.\ DP160102426.

\vspace{1 EM}

\noindent\textbf{Author Contributions}

\noindent  S.D.\ and H.M.W.\ developed the theory, S.D., S.S.\ and G.J.P.\ designed and performed the experiment. D.W.B.\ performed theoretical comparison of different measurement schemes. All the authors discussed the results and contributed to the writing of the manuscript. 

\vspace{1 EM}

\noindent\textbf{Additional Information}

\noindent Correspondence and requests for materials should be addressed to H.M.W.\ (\href{mailto:h.wiseman@griffith.edu.au}{h.wiseman@griffith.edu.au}) and G.J.P.\ (email: \href{mailto:g.pryde@griffith.edu.au}{g.pryde@griffith.edu.au}).\\
\vspace{1 EM}

\numberwithin{equation}{section}
\numberwithin{figure}{section}
\renewcommand{\figurename}{{\bf Appendix Figure}}
\renewcommand{\thefigure}{{\bf \arabic{figure}}}
\renewcommand{\theequation}{A\arabic{equation}}
\setcounter{equation}{0}
\setcounter{figure}{0}

\onecolumngrid

\appendix
\begin{figure}[h]
	\includegraphics{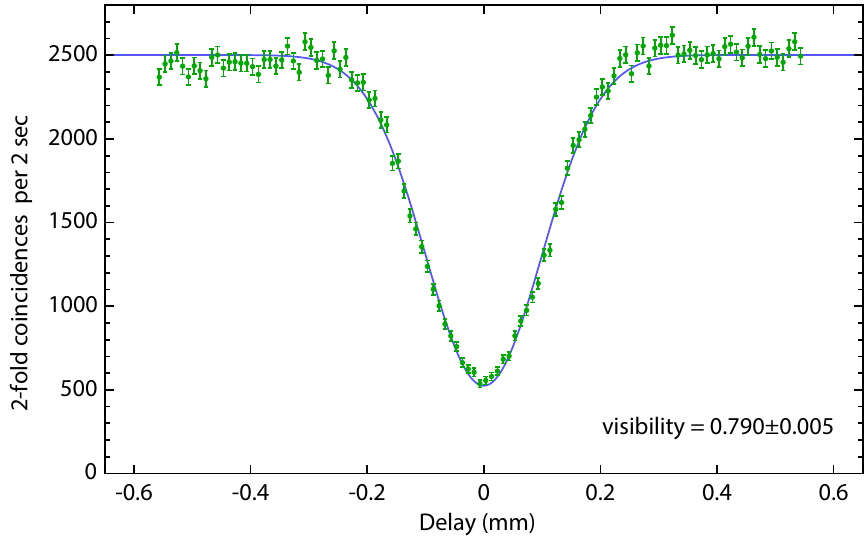}
	\caption{\label{fig:hom} {\bf Hong-Ou-Mandel interference at the CNOT gate}. Two-photon coincidence rate as a function of the photon indistinguishability, controlled by introducing a propagation delay in one of the photon's path. Maximum theoretical visibility is 0.8.}
\end{figure}
\section{Numerical simulation of the HPEA} \label{appnD} 
Consider the Heisenberg-limited interferometric phase estimation algorithm with $K+1=2$ (qubits) photons shown in Fig.~1d. Assume the input state is represented by $\hat{\rho}_{\rm in}$ which allows us to consider mixed input states. The state $\hat{\rho}^{(K)} \in {\mathbb B}_{2^{K+1}}$ (${\mathbb B}$ denoting Banach space) of the system before the first $X$-measurement on the $K$-th photon is
\beq\label{rho:Kth}
\hat{\rho}^{(K)} = \(\hat{U}^{2^K}\bigotimes_{k=1}^K \hat{I}\) \hat{\rho}_{\rm in}\,\(\hat{U}^{2^K}\bigotimes_{k=1}^K \hat{I}\)\dg,
\eeq
where 
\beq  \label{phase:mat}
U^p= 
\begin{pmatrix}
	1 & 0 \\
	0 & e^{ip\phi}
\end{pmatrix}, \qquad
{I} = 
\begin{pmatrix}
	1 & 0 \\
	0 & 1
\end{pmatrix}. 
\eeq 
Here $\phi$ is a random unknown phase in the interval $[0,2\pi)$ to be estimated. To find the outcome of the $X$-measurement on the $K$-th photon, we define the following two measurement operators
\beq \label{mmnt:opt:Kth}
\hat{M}_{r}^{(K)} = \hat{Q}_{r} \bigotimes_{k=1}^K \hat{I},  %\qquad {\rm for} \quad l = D, A.
\eeq 
where $r \in \{{\rm d}, {\rm a}\}$ is a measurement result, and $\hat{Q}_{r} = \op{r}{r}$ is the projection operator onto the $X$ basis of the $K$-th photon. Thus, the probability of finding the $K$-th photon in one of the $X$ eigenstates is 
\beq
P_{r}^{(K)} = \tr{\hat{\rho}^{(K)} \hat{M}_{r}^{(K)} {\hat{M}_{r}^{(K)}}\dg} = \tr{\hat{\rho}^{(K)} \hat{M}_{r}^{(K)}}.
\eeq
Whether the $K$-th qubit is found in $\ket{{\rm d}}$ or $\ket{{\rm a}}$ is determined by calling a random number and comparing it with the above probability. Depending on the outcome of this last step, the conditional system state $\hat{\rho}_r^{(K)}$ after the measurement on the $K$-th photon  is obtained according to \cite{WisMil10}
\beq
\hat{\rho}_r^{(K)} = {\cal J}\Big[\hat{M}_{r}^{(K)}\Big] \,\hat{\rho}^{(K)}/P_{r}^{(K)} = \hat{Q}_r \otimes \hat{\rho}_r^{(K-1)},
\eeq
where $\hat{\rho}_r^{(K-1)}  \in {\mathbb B}_{2^{K}}$ is the reduced state matrix of the other remaining $K$ photons. Here  ${\cal J}[\hat{O}]\hat{A} \equiv \hat{O} \hat{A}\hat{O}\dg$ for some arbitrary operators $\hat O$ and $\hat A$. To proceed with the next step of the protocol, the result of the previous measurement is used to decide whether feedforward should be applied or not. That is, the outcomes $r={\rm a}$ and $r={\rm d}$  correspond to control operation ON and OFF, respectively.  Therefore, in the reduced-dimension Hilbert space of the system, the state matrix before the measurement on the $(K-1)$-th photon when the feedback operation is ON can be expressed as 
\beq
\hat{\rho}^{(K-1)} = \hat{\mathcal{U}}^{(K-1)} \, \hat{\rho}_r^{(K-1)} \; {\hat{\mathcal{U}}^{(K-1)}}\dg,
\eeq
where 
\bqa  
\hat{\mathcal{U}}^{(K-1)} &\equiv& \left[\hat{U}^{2^{K-1}}\hat{\mathcal{R}}\left(\frac{\pi}{2}\right)\right] \bigotimes_{k=2}^K \hat{\mathcal{R}}\left(\frac{\pi}{2^k}\right), \\[4pt] 
{\mathcal{R}}(\theta) &=& 
\begin{pmatrix} 
	e^{i\theta} & 0 \\
	0 & 1
\end{pmatrix}, \label{FB:phase}
\eqa 
and if the feedforward is OFF,  the state matrix is 
\beq   \label{FB:off:K-1}
\hat{\rho}^{(K-1)} = \left(\hat{U}^{2^{K-1}}\bigotimes_{k=1}^K \hat{I}\right) \hat{\rho}_r^{(K-1)} \left(\hat{U}^{2^{K-1}}\bigotimes_{k=1}^K \hat{I}\right)\dg.
\eeq
Measurement on the $(K-1)$-th photon is described in the same way as that of the $K$-th photon. That is, by changing $K\to K-1$ we can use Eqns.~(\ref{mmnt:opt:Kth})-(\ref{FB:off:K-1}) to find the measurement result and the reduced state $\hat{\rho}_r^{(K-2)}  \in {\mathbb B}_{2^{K-1}}$ of the system. These steps are repeated for each qubit until the $0$-th one, for which, the measurement operator is simply the projector $\hat{M}_{r}^{(0)} = \hat{Q}_{r}$. Finally the same procedure as described in Methods is employed to calculate the Holevo variance. 
\section{Shot-noise limit analytical calculation} \label{appnE1} 
The asymptotic limit of the phase variance can be calculated in an interferometric phase estimation context to obtain the SNL which amounts to $V^{\rm SNL} \sim1/N$. This limit is valid when the number $N$ of resources goes to infinity. When $N$ is finite and small, as is the case here in our experiment, this relation does not hold at all. Instead, we were required to analytically calculate what is the SNL for small $N$'s \cite{BerWis09}. 

Consider the interferometer shown in Fig.~1a without multipassing, that is, $p=1$. Sending single-photon Fock state into one arm of the interferometer, the probability of detecting a photon in either of the output ports in an ideal experimental situation is given by
\beq  \label{prob:click}
P(u|\phi,\theta) = \frac{1}{2}\left[1+u\, {\rm cos}(\phi-\theta)\right],
\eeq
where $u \in \{-1,1\}$ labels the measurement outcome. Assuming that $N$ measurement results are obtained, we can represent them as a vector ${\bf u}_N = \(u_1, u_2, \cdots, u_N\)$ in which each $u_j$ is defined as above. Therefore, the probability for the sequence of measurement results is given by 
\beq \label{prob:u}
P({\bf u}_N|\phi,{{\bm \uptheta}}) = P(u_1|\phi,\theta_1)\, P(u_2|\phi,\theta_2)\; \cdots \;P(u_N|\phi,\theta_N),
\eeq
where the adjustable phase is varied according to $\theta_j = j {\pi}/{N}$. Now recall the Holevo variance in the phase estimate $V_H = \mu^{-2}-1$ where the sharpness $\mu = \abs{\an{e^{i\phi}}}$. Using the conditional probability given in Eq.~(\ref{prob:u}), $\mu$ can be written as \cite{BerWis09,BerBre01}
\beq \label{sharpness}
\mu = \frac{1}{2\pi} \sum_{{\bf u}_N} \abs{\int{e^{i\phi} P({\bf u}_N|\phi,{\bm \uptheta}) \,d\phi}}.
\eeq 
We can then calculate the Holevo variance for small $N$'s by solving this integral. For our experiment in which $N=3$, we could easily calculate the sharpness and find the exact standard quantum limit to be $V^{\rm SNL} =7/9$. However, as the number of resources increases this calculation gets complex and at some point even impossible to solve the integral exactly as the number of possible results goes up exponentially.

\section{Shot-noise limit experiment} \label{appnE2} 
To measure the SNL we have used the same experimental setup. Unentangled single photons were guided through the interferometer such that one of them was used as a probe system and the other one heralded the presence of the former. Instead of passing three photons once through the phase shift element, we sent three single photons sequentially one after the other and adjusted the controllable phase to $\theta_j$, respectively, for $j=1,2,$ and $3$. For each setting there would be two measurement outcomes $u_j$. This means for a fixed $\phi$ there are $2^3=8$ possible results (two of which, $\phi_1 = \phi_3 = \pm1$ and $\phi_2 = \mp1$, are not useful because they yield no information about the unknown phase). Let $n_{u_j}(\phi,\theta_j)$ represents the number of times that a particular outcome turns up out of an ensemble size $n_{\rm ens}=\sum_{u_j}n_{u_j}(\phi,\theta_j)$. Therefore, the probability of having the outcome ${\bf u}_3 = (u_1,u_2,u_3)$ for three independent measurement is
\beq
P({{\bf u}_3}|\phi) = \Pi_j \frac{n_{u_j}(\phi,\theta_j)}{n_{\rm ens}}, 
\eeq
and the true phase can be calculated using
\beq \label{phi:exp:SNL}
\phi = {\rm arg}  \sum_{{\bf u}_3}\int P({{\bf u}_3}|\phi) \, e^{i\phi}\,d\phi.
\eeq
On the other hand, $\phi_{\rm est}$ for three measurement outcomes ${\bf u}_3$ is
\beq \label{phi:the:SNL}
\phi_{\rm est}({\bf u}_3) = {\rm arg} \int \Pi_j\; P(u_j|\phi,\theta_j) \, e^{i\phi}\,d\phi.
\eeq
For a given $\phi$ we proceeded as the following to calculate the sharpness
\beq
\mu (\phi) = \abs{ \sum_{{\bf u}_3} \; P({{\bf u}_3}|\phi)\; e^{i(\phi-\phi_{\rm est})} }.
\eeq 
It is easy to work out the Holevo variance by averaging over $\phi$ in the same way as before.  
\section{Holevo variance for the protocols in Table II}\label{appnF}
For the case of adaptive measurements and entanglement but no multiple passes, we consider $3$ photons in a single spatio-temporal mode so they are indistinguishable. That is, by construction, the photons are identically prepared; the entanglement is symmetric under photon exchange.  
This single  mode could be over an extended time, so photons can be detected separately, and the controlled phase can be adjusted in between detections.
As discussed in the main text, if we had instead  considered three distinguishable photons in different modes, then having entanglement and adaptive measurements would be the most powerful scheme possible, and would give the same Holevo variance as using symmetric entanglement, multiple passes and adaptive measurements.

There are a total of 3 phases that need to be optimised.
Before the first detection, the controllable phase $\theta$ has no effect on the results.
This is because the system phase is averaged over, and it is only the relative phases that are important.
There are two possibilities for the first detection result, and values of the controllable phase $\theta$ need to be chosen for each.
There are four possibilities for the first two detection results, and again values of the controllable phase $\theta$ after those two detections need to be chosen.
This gives six phases, but changing the initial value of $\phi$ by $\pi$ reverses the significance of the detection results.
Because of this symmetry the number of phases that need be considered is reduced by a factor of $2$.
In addition, the entangled state needs to be optimised over.

Results for the case where the state is optimal for canonical measurements were given in Fig.~11 of Ref.~\cite{BerBre01}, where it was found that the phase variance was slightly above that for canonical measurements for $3$ photons.
That result shows that it is not possible to achieve the HL in that case, though it leaves open the possibility that slightly better performance (but still not at the HL) could be obtained by optimising over the state as well.
The result for that case was given in Fig.~6.10 of Ref.~\cite{BerryTh}, and the optimisation over the state gives a very slight improvement for $N=3$.
The exact value obtained was $0.5569202271898053$.

For the case with adaptive measurements and multiple passes but no entanglement, there are three general possibilities for $N=3$.
\begin{enumerate}
	\item One photon with three passes.
	\item One photon with two passes and one photons with one pass.
	\item Three single photons with a single pass each.
\end{enumerate}
The first is trivial because there is phase ambiguity so the Holevo variance is infinite.
The second can be treated using the approach of Sec.~IV of Ref.~\onlinecite{BerWis09}, where an equivalent two-mode state in a single time mode is considered.
There it was found that the ideal canonical measurement gives a variance of $2/N+1/N^2$ (see Eq.~(4.4)).
For $N=3$ this gives $7/9= 0.777\ldots$
A result for the third possibility was given in Fig.~6.7 of Ref.~\onlinecite{BerryTh}, though with a restricted optimisation of the adaptive measurements, and obtained a Holevo variance of $0.5609756097560981$.
We have recalculated the variance with full optimisation over $\theta$, and found that the variance is unchanged.

Finally we consider the case of symmetric entanglement and multiple passes, but no adaptive measurements.
There are three possibilities again, and again the case with one photon and three passes is trivial.
For the others, it is necessary to optimise over the controlled phases $\theta$ and the state.
The optimisation over the phases is simpler than for the adaptive case, because the phase does not depend on the detection results.
It was found that for one photon with two passes and another with a single pass the minimum Holevo variance was $2$.
The best result was for three entangled photons in a single mode and single passes, in which case the minimum Holevo variance was $0.6546809936433506$. We note that if one were to drop the requirement  of symmetry on the entangled state, one could obtain a slightly smaller variance of $0.6054864794870138$, using an entangled state across three modes.  

These calculations were performed in the following way.
First, a state of three successive photons in different spatio-temporal modes can be given as
\begin{equation}
\ket{\psi} = \sum_{j,k,l\in\{0,1\}} \psi_{j,k,l} \ket{j,k,l},
\end{equation}
where $j$, $k$, and $l$ indicate which polarisation each photon is in.
This formalism can also be used to treat multiple photons in the same spatio-temporal mode, by using a symmetric state.

Then the operation of measuring a photon as being in one polarisation or the other on the first mode can be represented by
\begin{equation}
\ip{a}{\psi} = \sum_{j,k,l\in\{0,1\}} \psi_{j,k,l} \ip{a}{j}\ket{k,l},
\end{equation}
where
\begin{equation}
\ket{a} = \frac 1{\sqrt{2}}(\ket{0}+(-1)^a\ket{1}),
\end{equation}
for $a\in\{0,1\}$.
It is easy to see that
\begin{equation}
\ip{a}{j}=\frac 1{\sqrt{2}} (-1)^{aj}.
\end{equation}
For three measurement results, $a$, $b$, and $c$, the inner product we need is
\begin{align}
\ip{a,b,c}{\psi} &= \sum_{j,k,l\in\{0,1\}} \psi_{j,k,l} \ip{a}{j}\ip{b}{k}\ip{c}{l}\nonumber \\
&=\sum_{j,k,l\in\{0,1\}} \psi_{j,k,l} (-1)^{aj+bk+cl}
\end{align}
Similarly to how described in previous sections, the probability of obtaining the measurement result is given by the absolute value squared:
\begin{equation}
P(a,b,c)=|\ip{a,b,c}{\psi}|^2.
\end{equation}
Next we explain how to take the phase into consideration.
Without loss of generality we can take the first controlled phase $\theta_0$ to be zero (since we average over $\phi$).
The second controlled phase $\theta_1$ can depend on $a$, and the third $\theta_2$ can depend on $a$ and $b$.
The change in the state with these controlled phases and the system phase $\phi$ is
\begin{equation}
\ket{\psi(\phi)} = \sum_{j,k,l\in\{0,1\}} e^{i[j\phi+k(\phi-\theta_1)+l(\phi-\theta_2)]}\psi_{j,k,l} \ket{j,k,l}.
\end{equation}
This state is convenient to use for calculation, but will not correspond to the physical state at any stage.
In reality the first photon would be detected before the phase $\phi-\theta_1$ is applied to the second photon, and so forth.
The important quantity is the inner product
\begin{equation}
\ip{a,b,c}{\psi(\phi)} =\sum_{j,k,l\in\{0,1\}} e^{i[j\phi+k(\phi-\theta_1)+l(\phi-\theta_2)]} (-1)^{aj+bk+cl}\psi_{j,k,l}
\end{equation}
which enables us to calculate the probability as a function of $\phi$
\begin{equation}
P(a,b,c|\phi)=|\ip{a,b,c}{\psi(\phi)}|^2.
\end{equation}
Next, we determine the Holevo variance with sharpness $\mu$ given by
\begin{equation}
\mu = \frac 1{2\pi}\sum_{a,b,c\in\{0,1\}} \left| \int \, e^{i\phi} P(a,b,c|\phi) d\phi \right|.
\end{equation}
We can calculate the integral as
\begin{equation}
\int \, e^{i\phi} P(a,b,c|\phi)d\phi = \sum_{j+k+l+1=j'+k'+l'} e^{i[(j-j')\phi+(k-k')(\phi-\theta_1)+(l-l')(\phi-\theta_2)]} (-1)^{a(j-j')+b(k-k')+c(l-l')}\psi_{j,k,l}\psi^*_{j,k,l}.
\end{equation}
This formula can be used to minimise the phase variance with various types of measurement.
If the measurement is allowed to be adaptive, then $\theta_1$ can depend on $a$ and $\theta_2$ can depend on $a$ and $b$.
If it is not adaptive, then $\theta_1$ and $\theta_2$ would need to be chosen independently of the measurement results.
Restrictions on the state can also be imposed, for example by requiring it to be separable between the three modes, or by requiring it to be symmetric between the three photons to correspond to three photons in the one mode.

\end{document}